\documentclass[conference]{IEEEtran}
\IEEEoverridecommandlockouts

\usepackage{cite}
\usepackage{amsmath,amssymb,amsfonts}
\usepackage{algorithmic}
\usepackage{graphicx}
\usepackage{textcomp}
\usepackage{xcolor}
\usepackage{multirow}
\usepackage{makecell}
\def\BibTeX{{\rm B\kern-.05em{\sc i\kern-.025em b}\kern-.08em
    T\kern-.1667em\lower.7ex\hbox{E}\kern-.125emX}}
\begin{document}


\title{Subject-Independent Imagined Speech Detection via Cross-Subject Generalization and Calibration
\thanks{This work was partly supported by Institute of Information \& Communications Technology Planning \& Evaluation (IITP) grant funded by the Korea government (MSIT) No. RS--2021--II212068, Artificial Intelligence Innovation Hub, No. RS--2024--00336673, AI Technology for Interactive Communication of Language Impaired Individuals, and No. RS--2019--II190079, Artificial Intelligence Graduate School Program (Korea University)).}}

\author{
\IEEEauthorblockN{Byung-Kwan Ko}
\IEEEauthorblockA{\textit{Dept. of Artificial Intelligence} \\
\textit{Korea University}\\
Seoul, Republic of Korea \\
leaderbk525@korea.ac.kr} \\
\and

\IEEEauthorblockN{Soowon Kim}
\IEEEauthorblockA{\textit{Dept. of Artificial Intelligence} \\
\textit{Korea University} \\
Seoul, Republic of Korea \\
soowon\_kim@korea.ac.kr}\\
\and

\IEEEauthorblockN{Seo-Hyun Lee}
\IEEEauthorblockA{\textit{Dept. of Brain and Cognitive Engineering} \\
\textit{Korea University} \\
Seoul, Republic of Korea \\
seohyunlee@korea.ac.kr}\\
}

\maketitle

\begin{abstract}
Achieving robust generalization across individuals remains a major challenge in electroencephalogram-based imagined-speech decoding due to substantial variability in neural activity patterns.
This study examined how training dynamics and lightweight subject-specific adaptation influence cross-subject performance in a neural decoding framework.
A cyclic inter-subject training approach, involving shorter per-subject training segments and frequent alternation among subjects, led to modest yet consistent improvements in decoding performance across unseen target data.
Furthermore, under the subject-calibrated leave-one-subject-out scheme, incorporating only 10~\% of the target subject’s data for calibration achieved an accuracy of 0.781 and an AUC of 0.801, demonstrating the effectiveness of few-shot adaptation.
These findings suggest that integrating cyclic training with minimal calibration provides a simple and effective strategy for developing scalable, user-adaptive brain–computer interface systems that balance generalization and personalization.
\end{abstract}

\begin{IEEEkeywords} 
brain-computer interface, deep learning, generalization, imagined speech, subject-independent;
\end{IEEEkeywords}
  
\section{INTRODUCTION}
Brain–computer interfaces (BCIs) have emerged as promising assistive and rehabilitative technologies that enable communication and functional recovery for individuals with severe motor or speech impairments.~\cite{chaudhary2016brain, prabhakar2020framework} Among various paradigms~\cite{townsend2004continuous, lee2020continuous, 8937838}, imagined speech BCIs aim to decode internally generated linguistic representations directly from neural activity, offering a natural and intuitive communication pathway without the need of overt articulation~\cite{wang2013analysis, bulthoff2003biologically}. Non-invasive electroencephalogram (EEG) has been widely adopted for this purpose due to its high temporal resolution and clinical accessibility~\cite{ding2013changes, panwar2020modeling}. Recent progress in deep learning-based neural decoding has further enhanced the identification of speech-related brain activity with increasing precision~\cite{lee2003pattern, kamble2022deep, de2024imagined}. Despite these advances, decoding internally generated speech remains challenging topics in neural signal processing, due to the subtle and overlapping spatiotemporal features involved in silent linguistic imagery.


Most existing studies on imagined-speech decoding have been conducted in subject-dependent settings, where models are trained and tested on data from the same individual~\cite{mugler2014direct, suk2011subject, song2017novel, lee1996multiresolution}. This approach enables the network to capture user-specific EEG characteristics and often yields strong performance under controlled conditions~\cite{ramsey2018decoding}. Our previous work~\cite{ko2025imagined} proposed a multi-receptive field EEGNet (MRF-EEGNet) designed to enhance asynchronous detection of imagined-speech and idle states. While this architecture and method demonstrated strong within-subject performance and robustness to temporal variability, it was evaluated exclusively under a subject-dependent cross-validation scheme. Consequently, its generalization ability across unseen subjects remained unexplored~\cite{lee1999integrated}.

Achieving subject-independent generalization in EEG-based BCIs remains a continuing challenge~\cite{waytowich2016spectral, 10411116}. EEG signals exhibit substantial inter-subject variability, stemming from differences in scalp conductivity, cortical geometry, electrode placement, and individual neural strategies during task execution~\cite{kim2015abstract, ray2015subject}. Furthermore, EEG is inherently non-stationary, with signal distributions fluctuating across sessions due to cognitive fatigue, attention, and environmental factors~\cite{ wang2013analysis}. These physiological and contextual variations lead to distribution shifts between subjects, posing a challenge to learning a unified neural representation space~\cite{angrick2021real}. Although deep neural networks have advanced the field through hierarchical feature extraction~\cite{lee2018deep, cho2021neurograsp, 9991178,  lee2015motion, lee1995multilayer, schirrmeister2017deep}, they still face difficulties in capturing subject-invariant representations when training data are limited and heterogeneous. As a result, models trained on one group of users tend to show limited generalization to new individuals, highlighting the ongoing gap between controlled experimental performance and practical usability.


\begin{figure*}[t]
\centering
    \includegraphics[width=\textwidth]{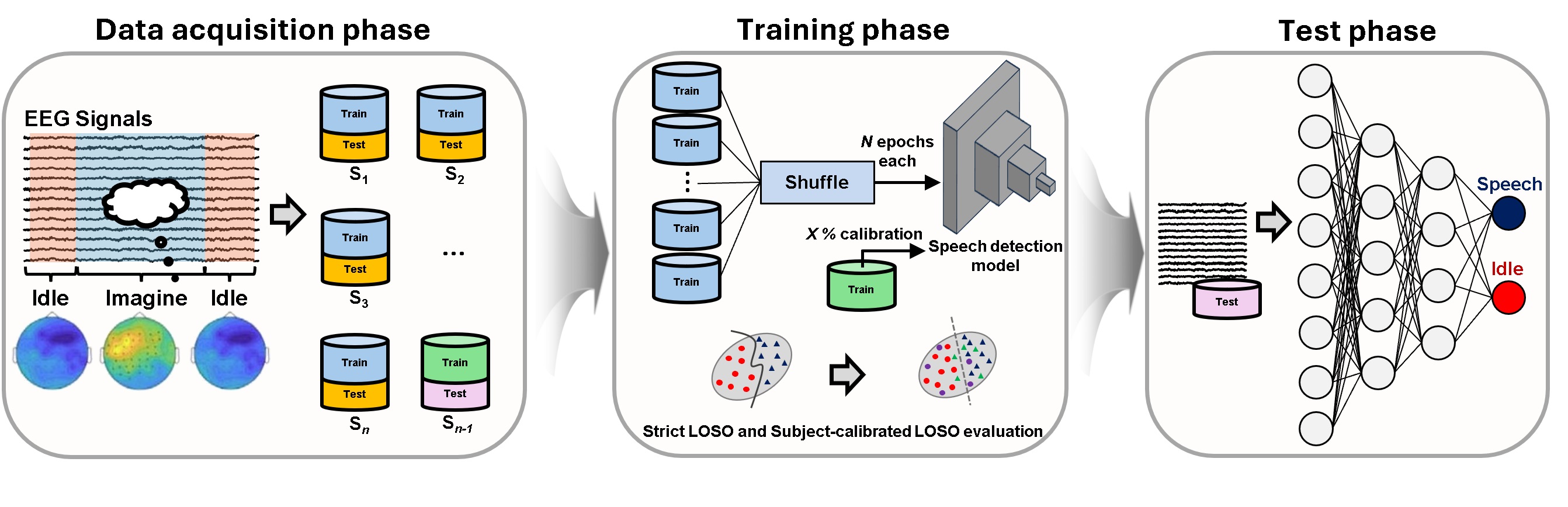}
    \caption{Overall experimental pipeline comprising the data acquisition, training, and test phases under two evaluation schemes: strict LOSO and subject-calibrated LOSO. In the strict LOSO setup, the model is trained using data from all subjects except \( S_{n-1} \)and directly evaluated on the unseen subject. In the subject-calibrated LOSO scheme, a small portion of \( S_{n-1} \)'s data is used for fine-tuning before testing, leading to final classification outputs between \textit{speech} and \textit{idle} states.}
    \label{figure1}    
    \vspace{-0.3cm}
\end{figure*}

The present study extends our previous MRF-CNN framework toward cross-subject generalization in imagined-speech detection. We systematically investigate two complementary evaluation schemes:
(1) a strict leave-one-subject-out (LOSO) approach, which assesses the model’s ability to generalize without any exposure to the target subject’s data; and
(2) a subject-calibrated LOSO (SC-LOSO) validation, where a small portion of the target subject’s data is used for adaptation prior to testing.
Through these analyses, we aim to quantify both the degree of inter-subject variability and the potential benefits of minimal calibration for enhancing cross-user transferability.
The findings contribute to a deeper understanding of the generalization challenges in subject-independent EEG decoding and offer preliminary evidence that lightweight calibration can mitigate inter-subject discrepancies, suggesting practical directions for future adaptive imagined-speech BCI research.

\section{MATERIALS AND METHODS}

\subsection{Data and Model Architecture}
The dataset used in this study was obtained from our previous works~\cite{ko2025imagined}.
EEG data were collected from 6 subjects performing imagined speech tasks.
Each trial was segmented into 500-ms windows (shift size of 50-ms), and the segments were temporally aligned using an overt-speech-based ground-truth labeling method~\cite{soroush2023nested}.
The same preprocessing pipeline was applied as described in our previous study~\cite{ko2025imagined}, including a 5th-order bandpass filter (0.5–125 Hz), notch filters at 60 Hz and 120 Hz, and common average re-referencing. Labels were assigned as 0 for the idle state and 1 for the speech state.

The model was based on MRF-EEGNet~\cite{ko2025imagined}. The architecture integrates multi-receptive-field~\cite{kong2020hifi} convolutional layers and a long short-term memory module~\cite{lee1997new} to capture both multi-scale temporal dynamics and sequential dependencies in EEG signals~\cite{si2021imagined}.
All the parameters and kernel settings were kept identical to the original implementation to ensure that any performance differences in this study arise solely from the evaluation scheme rather than architectural modifications.\\

\subsection{Experiment Method}
To evaluate cross-subject generalization, two LOSO schemes were implemented as shown in Fig.~\ref{figure1}.

In the strict LOSO validation, the model was trained solely on the training data of all subjects except the target subject, and tested on the held-out subject’s test data without any exposure to that subject’s EEG during training~\cite{lawhern2018eegnet}.
This setting represents a fully subject-independent evaluation and quantifies the model’s inherent generalization ability to unseen subjects~\cite{ray2015subject}. Each model was trained for $N$ epochs using the Adam optimizer with a learning rate of $1\times10^{-3}$ and a batch size of 64, while all other parameters followed the original MRF-EEGNet implementation.

On the other hand, the SC-LOSO scheme simulates a practical scenario where a small portion of the target subject’s data is available for lightweight calibration.
In this setting, the pretrained model from the strict LOSO phase was fine-tuned for two epochs using a learning rate of $5\times10^{-3}$ and a batch size of 64.
Only a subset ($X~\%$, $X\in{5,10,15}$) of the target subject’s training samples was used for calibration, and the remaining unseen samples were reserved for testing.
This setup was designed to evaluate the trade-off between calibration effort and cross-subject transferability.

\begin{table*}[t]
\centering
\caption{Performance comparison between subject-dependent and subject-independent evaluation schemes.}
\label{table1}
\renewcommand{\arraystretch}{1.15}
\setlength{\tabcolsep}{6pt}
\begin{tabular*}{\textwidth}{@{\extracolsep{\fill}} c c c c c c c c @{}}
\hline
\textbf{Evaluation Type} & \textbf{Scheme} & \textbf{Calibration (\%)} & \textbf{Train Config. ($N, R$)} & \textbf{Accuracy} & \textbf{F1-score} & \textbf{MCC~\cite{chicco2020advantages}} & \textbf{AUC} \\
\hline
\textbf{Subject-Dependent} & -- & -- & -- & {0.817 ± 0.057} & {0.777 ± 0.060} & {0.561 ± 0.117} & {0.860 ± 0.048} \\
\hline

\multirow{8}{*}{\centering\arraybackslash\textbf{Subject-Independent}} 
 & \multirow{2}{*}{\centering\arraybackslash\textbf{Strict LOSO}} & -- & (5, 6) & 0.694 ± 0.036 & 0.503 ± 0.085 & 0.044 ± 0.143 & 0.562 ± 0.114 \\
 &                      & -- & (3, 10) & 0.670 ± 0.092 & 0.542 ± 0.037 & 0.123 ± 0.069 & 0.619 ± 0.067 \\
\cline{2-8}
 & \multirow{6}{*}{\centering\arraybackslash\textbf{SC-LOSO}} 
 & 5 & (5, 6) & 0.758 ± 0.056 & 0.626 ± 0.123 & 0.310 ± 0.191 & 0.750 ± 0.102 \\
 &                  & 5 & (3, 10) & 0.755 ± 0.060 & 0.658 ± 0.091 & 0.351 ± 0.155 & 0.780 ± 0.080 \\
 \cline{3-8}
 &                  & 10 & (5, 6) & 0.778 ± 0.059 & 0.651 ± 0.140 & 0.358 ± 0.791 & 0.791 ± 0.101 \\
 &                  & 10 & (3, 10) & 0.781 ± 0.053 & 0.666 ± 0.109 & 0.382 ± 0.162 & 0.801 ± 0.078 \\
 \cline{3-8}
 &                  & 15 & (5, 6) & 0.777 ± 0.054 & 0.690 ± 0.075 & 0.400 ± 0.134 & 0.808 ± 0.081 \\
 &                  & 15 & (3, 10) & 0.785 ± 0.054 & 0.698 ± 0.079 & 0.406 ± 0.150 & 0.809 ± 0.071 \\
\hline
\end{tabular*}
\end{table*}

A cyclic subject-stream training strategy was employed, in which the training subjects were iteratively shuffled at the beginning of each cycle and each subject’s data were trained for $N$ epochs per cycle. 
Two configurations were designed to analyze the trade-off between rotation frequency and per-subject training depth: a \textit{frequent-rotation} setup ($N{=}3$ and $R{=}10$), 
where subjects change often with shallow updates per cycle, and a \textit{deep-subject} setup ($N{=}5$ and $R{=}6$), where each subject undergoes longer updates before the next rotation. 
These settings jointly allow examination of how training dynamics and subject-switching frequency 
affect cross-subject generalization. 

\section{RESULTS AND DISCUSSION}

Table \ref{table1} summarizes a detailed comparison of decoding performance across different evaluation schemes with metrics of accuracy, F1-score, matthews correlation coefficient (MCC)~\cite{chicco2020advantages}, and AUC.
As expected, the subject-dependent configuration yielded the highest accuracy (0.817 ± 0.057) and AUC (0.860 ± 0.048), representing an empirical upper bound for subject-specific optimization.
Under the strict LOSO condition, however, the overall performance dropped markedly, confirming the difficulty of cross-subject generalization~\cite{suk2014predicting} in EEG-based imagined-speech decoding.

Interestingly, when the training configuration changed from ($N{=}5$ and $R{=}6$) to ($N{=}3$ and $R{=}10$)—that is, using shorter per-subject epochs but more training cycles—the model exhibited slightly improved performance.
This modest gain suggests that the model became somewhat more effective at detecting speech-related brain states rather than overfitting to idle states.
Increasing the number of randomized inter-subject cycles exposed the network to a wider range of speech-related EEG patterns, improving its ability to extract invariant features of the speech state.
In contrast, longer per-subject epochs tended to emphasize subject-specific temporal dynamics, which were less transferable across individuals. Thus, the cyclic configuration with frequent subject alternations acted as an implicit regularizer, reducing overfitting and stabilizing optimization through more diverse gradient updates.

Under the SC-LOSO condition, all performance metrics improved progressively with increasing calibration ratio.
In particular, incorporating only 5~\% of calibration data ($N{=}3$ and $R{=}10$) increased accuracy from 0.670 to 0.755 and AUC from 0.619 to 0.780, while 10~\% calibration further raised them to 0.781 and 0.801, respectively. Beyond 10–15~\%, the gains became marginal, suggesting a saturation effect. These results highlight that even minimal subject-specific calibration ($\leq$10~\%) can substantially compensate for inter-individual variability, bridging much of the performance gap between strict LOSO and fully subject-dependent settings.

\begin{figure}[ht]
    \centering
    \includegraphics[width=0.9\columnwidth]{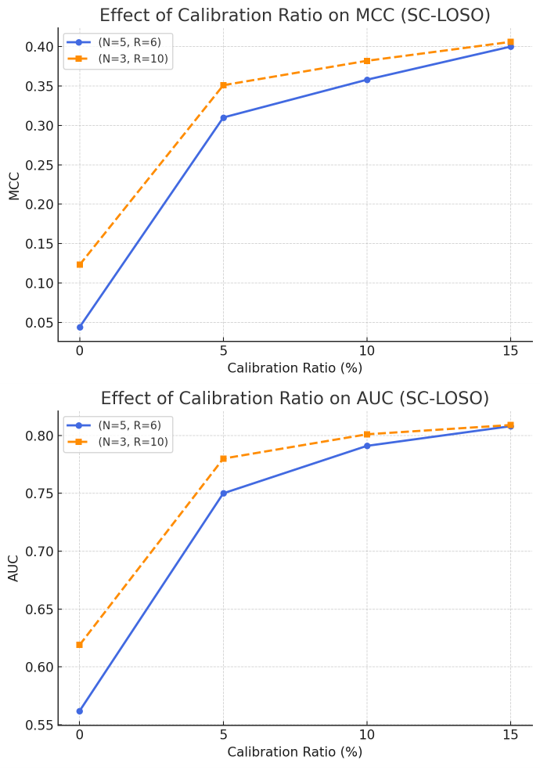}
    \caption{Comparison of MCC and AUC scores under the SC-LOSO scheme across different calibration ratios (5 \%, 10 \%, and 15 \%). The blue solid line represents the training configuration of ($N{=}5$ and $R{=}6$), and the orange dashed line corresponds to ($N{=}5$ and $R{=}6$).}
    \label{figure2}
    \vspace{-0.3cm} 
\end{figure}

Collectively, the findings underscore two complementary strategies for enhancing generalization in imagined-speech BCIs:
(1) employing shorter but more diverse inter-subject training cycles to learn invariant neural representations associated with the speech state, and
(2) applying lightweight few-shot calibration to align the learned feature space with the target subject’s neural characteristics.
This cyclic training paradigm also acts as an implicit regularizer, mitigating overfitting to individual EEG distributions while stabilizing optimization through diverse gradient updates.
The resulting feature alignment from minimal calibration demonstrates that even a small amount of user-specific adaptation can substantially enhance cross-subject transferability, highlighting a path toward practical and subject-scalable imagined-speech BCIs that balance generalization and personalization.

\section{CONCLUSIONS}
This study explored subject-independent imagined-speech decoding using MRF-EEGNet by systematically varying the training configuration and incorporating subject-specific calibration.
Through a series of LOSO experiments, we examined how the training dynamics and limited calibration data influence cross-subject generalization.
The results showed that the cyclic training configuration produced modest but consistent gains in cross-subject generalization, with similar effects observed during the calibration phase.
This improvement suggests that cyclic exposure to multiple subjects helps the model extract invariant speech-related EEG features and mitigates overfitting to subject-specific temporal patterns.
In addition, few-shot calibration under the SC-LOSO scheme demonstrated that even a small amount ($\leq$10~\%) of target-subject data can effectively adjust the learned representation, substantially narrowing the performance gap between subject-dependent and independent decoding.
Overall, these findings indicate that training dynamics and lightweight adaptation play critical roles in enhancing the robustness and generalizability of imagined-speech BCIs, offering practical insights for designing user-independent yet personalized decoding systems. Future work will focus on expanding the dataset and exploring data augmentation strategies to further assess and improve subject-independent performance~\cite{kaifosh2025generic}.


\bibliography{References}

\bibliographystyle{IEEEtran}


\end{document}